\documentclass[%
apl
 sd,%
 amsmath,amssymb,
 reprint,
]{revtex4-1}

\usepackage{graphicx}
\usepackage{dcolumn}
\usepackage{bm}
\usepackage[caption=false,font=normalsize,labelfont=sf,textfont=sf]{subfig}
\begin{document}


\title{Thickness Engineered Tunnel Field-Effect Transistors based on Phosphorene}

\author{Fan W. Chen}
\email{fanchen@purdue.edu}

\author{Hesameddin Ilatikhameneh}%

\author{ Tarek A. Ameen}


\author{ Gerhard Klimeck}

\author{ Rajib Rahman}
%

\affiliation{ Network for Computational Nanotechnology (NCN), Purdue University, West Lafayette, IN 47906, USA.\\}

\begin{abstract}
Thickness engineered tunneling field-effect transistors (TE-TFET) as a high performance ultra-scaled steep transistor is proposed. This device exploits a specific property of 2D materials: layer thickness dependent energy bandgap (Eg). Unlike the conventional hetero-junction TFETs, TE-TFET uses spatially varying layer thickness to form a hetero-junction. This offers advantages by avoiding the interface states and lattice mismatch problems. Furthermore, it boosts the ON-current to 1280$\mu A/\mu m$ for 15nm channel length. TE-TFET shows a channel length scalability down to 9nm with constant field scaling  $E = V_{DD}/L_{ch}= 30V/nm$. Providing a higher ON current, phosphorene TE-TFET outperforms the homojunction phosphorene TFET and the TMD TFET in terms of extrinsic energy-delay product. In this work, the operation principles of TE-TFET and its performance sensitivity to the design parameters are investigated by the means of full-band atomistic quantum transport simulation. 
\end{abstract}

\keywords{Tunnel Field Effect Transistor (TFET), Phosphorene, Device Scaling, Quantum Transport, NEGF, Heterojunction}
\maketitle

\section{Introduction}

{S}{ince} first experimental realization of $SS < 60mV/dec$\cite{appenzeller2004band} in tunnel field-effect transistors (TFETs), these devices have been the main candidate for reduction of supply voltage $V_{DD}$ and energy consumption in electronic devices. TFETs lower the energy consumption of a transistor by removing the hot carrier injection from the source region of the transistor. However, TFETs have 2 main challenges: 1) small ON-current and 2) channel length scaling. 

The small ON-current challenge of TFET is even more pronounced in the conventional CMOS channel materials, namely Si and Ge. These materials have an indirect gap which requires phonon assistance for band-to-band tunneling (BTBT). Si has also a large Eg which further reduces ${I_{ON}}$. Smaller band gap channel materials such as Ge can improve the tunneling current and ${I_{ON}}$, however they also increase the ${I_{OFF}}$, hence degrade the $I_{ON}/I_{OFF}$ ratio\cite{toh2008device}. Previously, several designs have been proposed to increase the ON-current of TFETs such as 1) heterostructure channels \cite{avci2013heterojunction, shih2011sub}, 2) dielectric engineering \cite{1, ilatikhameneh2016design} 3) internal polarization \cite{li2015polarization}, and 4) 2D materials \cite{ilatikhameneh2015tunnel, chen2016configurable, 7301966}. 

\begin{figure}[!t]
\centering
\subfloat[]{\includegraphics[width=0.34\textwidth]{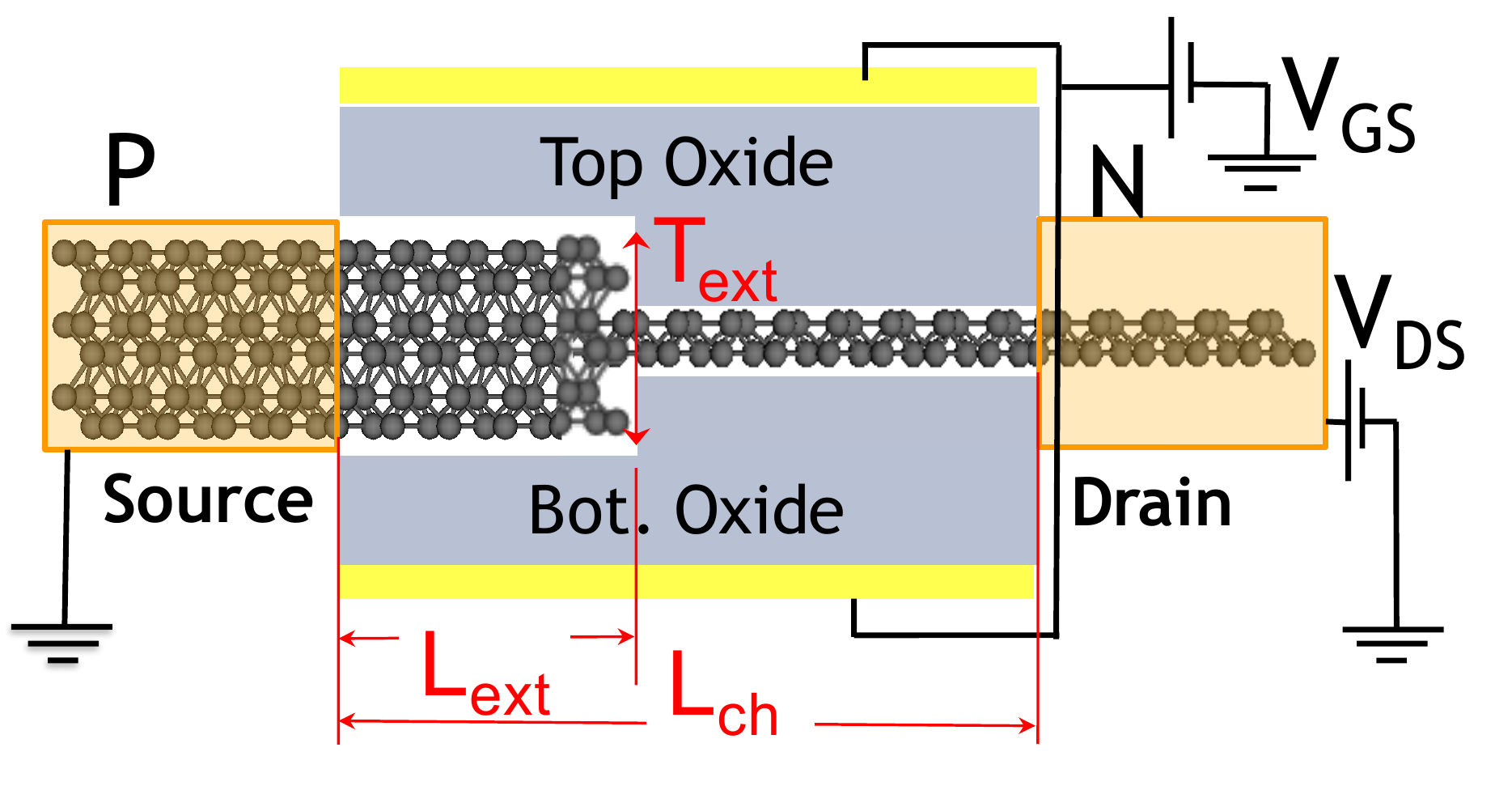}%
\label{1a}}
\vfil
\subfloat[]{\includegraphics[width=0.221\textwidth]{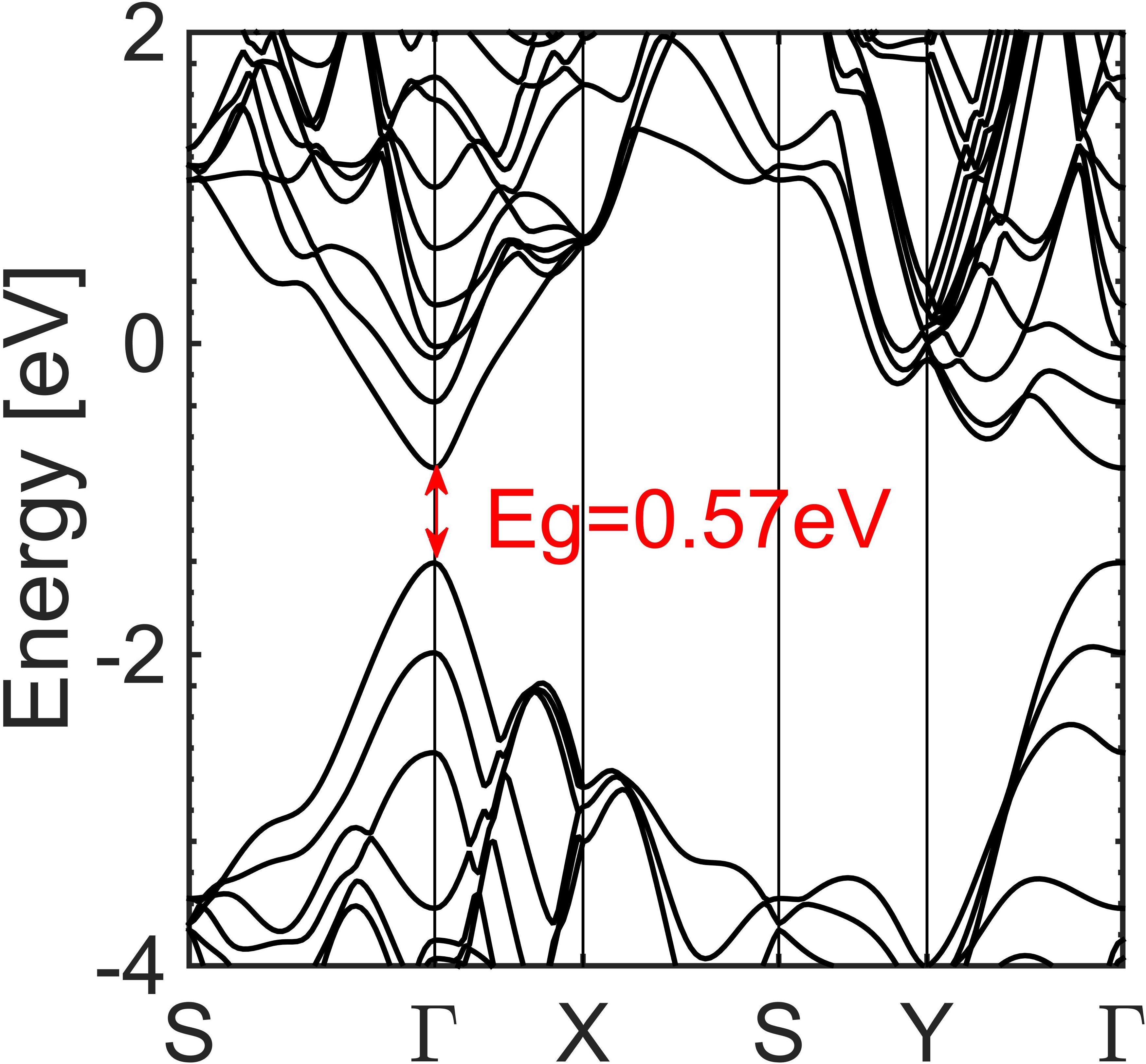}%
\label{1b}}
\hfil
\subfloat[]{\includegraphics[width=0.22\textwidth]{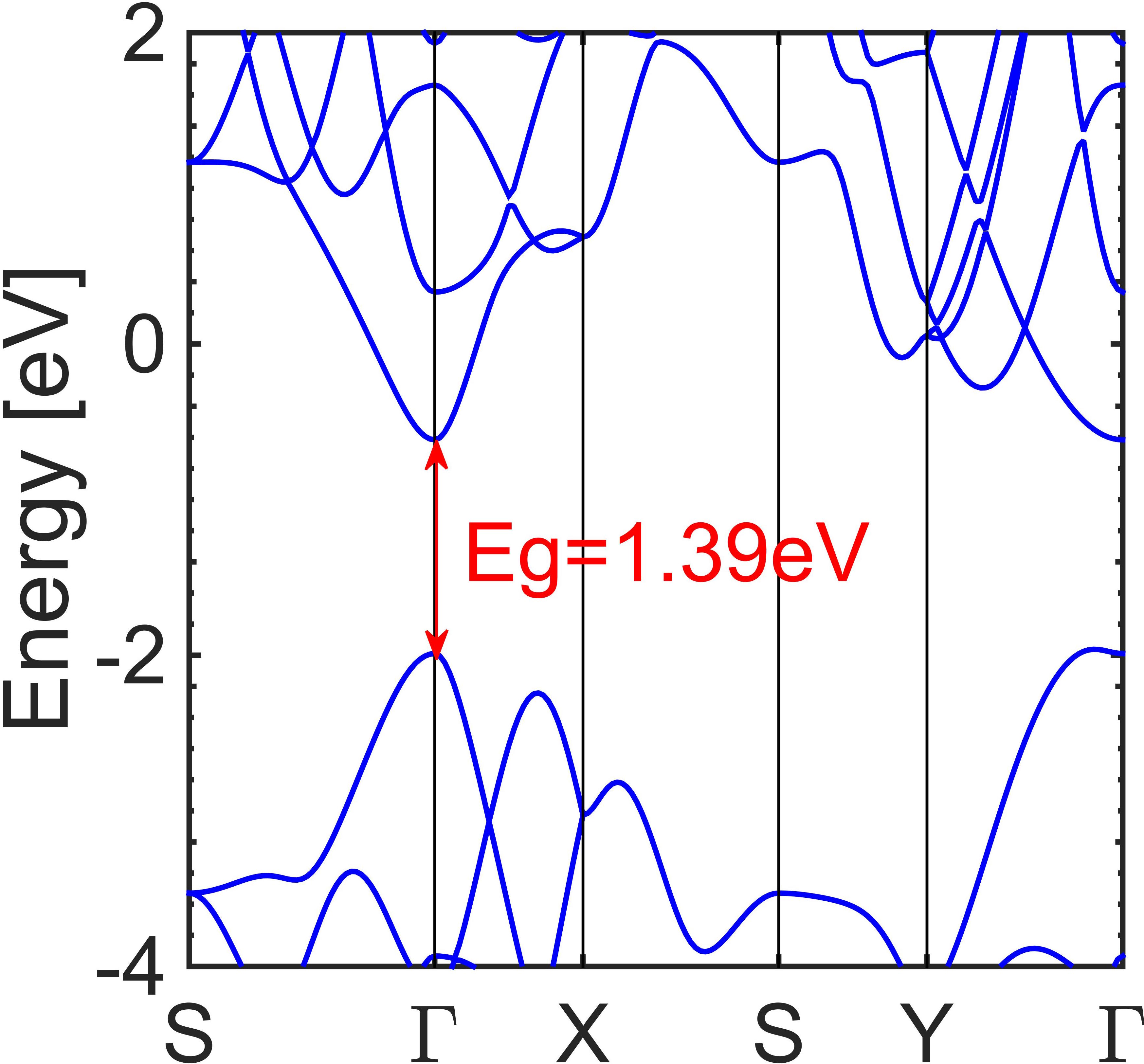}%
\label{1c}}
\caption{(a) The device structure of layer engineered TFET (TE-TFET) based on phosphorene. The device has a small Eg in the source and the channel region near the source and a large Eg region in the rest of the device. The layer thickness and the length of the small band gap region inside the channel are denoted by $L_{ext}$ and $T_{ext}$. The band structure of (b) 3L phosphorene with Eg = 0.57eV and (c) 1L phosphorene with Eg = 1.39eV. }
\label{1}
\end{figure}

Heterostructure channels improve the performance of TFETs by using small Eg as source, Si as channel material to improve ${I_{ON}}$ while keeping ${I_{OFF}}$ small. Unfortunately, the large lattice mismatch \cite{SiGe1, vandooren2013analysis} and interface states \cite{kim2015ge, 10nm, 5611586} between the materials prevent the formation of an ideal heterojunction. Artifical heterojunctions based on a single channel material have been achieved in graphene by varying the width of graphene nanoribbon (GNR)\cite{lam2010simulation}. However, the edge roughness and device-to-device variations due to the lack of atomic level control in top down fabrication pose a big challenge for their technology development\cite{luisier2009performance, yoon2007effect, basu2008effect}.

Novel 2D materials have interesting properties which can be used to provide artificial heterostructures. The bandgap of transition metal dichacogenides (TMDs), graphene and phosphorene depends on the layer thickness \cite{Chu2015bilayerMoS2, thicknessdependentTMD, cai2014layer, 24, chen2015transport}. In these materials, artificial heterojunctions can be achieved by spatially varying the layer thickness\cite {howell2015investigation}. Unlike the GNR heterojunctions where a sub-nanometer width control is required, a spatially varying layer thickness can be easily achieved with 2D material exfoliation and transfer techniques \cite{mas20112d, coleman2011two} . Therefore, a thickness engineered TFET (TE-TFET) which exploits this spatially varying layer thickness technique is proposed in this letter. TE-TFET is designed to have a small Eg in source and the channel near source and larger Eg in the rest of device. TE-TFET can be applied to any material that has a band gap dependence on layer thickness. In this work, phosphorene is chosen as the channel material due to the fact that multi-layer phosphorene is direct gap material \cite{tran2014layer, 6905726} and its bandgap range includes the optimum bandgap of 1.2 $q V_{dd}$ for TFET applications \cite{7331599, ameen2015few, liu2015device}. Although, the bandgap of TMD flakes depends on the flake thickness, only monolayer TMDs are direct gap materials.

  A TE-TFET has several advantages: (1) Artificial heterojunction structure avoids the problems with lattice mismatch and interface states observed in a conventional heterojunction; (2) the ON-state current can be enhanced due to the small bandgap and small tunnel distance in source-channel interface; (3) the OFF-state current remains small because of the large Eg barrier inside the channel. 
   
   The device structure of the TE-TFET based on phosphorene is shown in Fig.~\ref{1}a. The layer thickness and the length of the small band gap region inside the channel region are denoted by $L_{ext}$ and $T_{ext}$. The dependence of device performance on these design parameters and total channel length $L_{ch}$ will be discussed in details in section III. Finally, the capacitance voltage (CV) characteristics and energy delay product comparison with homojunction phosphorene TFET are discussed.

\section{Simulation Details}
\begin{table}[!t]
\renewcommand{\arraystretch}{1.3}
\caption{Phosphorene Parameters}
\label{tableI}
\centering

\begin{tabular}{|c|c|c|c|c|}
\hline
Layer & 1 & 2 & 3 & 4\\
\hline
Eg(eV) & 1.390 &0.803 & 0.570& 0.481\\
\hline
$\epsilon^{in}$ & 4.56 &7.41 & 8.77& 9.98\\
\hline
$\epsilon^{out}$ & 1.36 &1.52 &1.80& 2.04\\
\hline
\end{tabular}
\end{table}
The Hamiltonian of phosphorene is represented using a 10 band $sp^3d^5s^*$ second nearest neighbor tight binding model.  The tight-binding parameters are well calibrated to match the band structure and effective mass from density function theory (DFT) HSE06 by the standard mapping method  \cite{tan2015tight, tan2013empirical}. Fig.~\ref{1}b and Fig.~\ref{1}c show the tight-binding bandstructure of 3L- and 1L-phosphorene, respectively. The Eg of phosphorene flakes with different number of layers are listed in Table~\ref{tableI}. The relative permittivity for both in-plane $\epsilon^{in}$ and out-of-plane $\epsilon^{out}$ are taken from \cite{wang2015native} and are also listed in Table~\ref{tableI}. All the transport characteristics of the TE-TFET have been simulated using the self-consistent Poisson-Non Equilibrium Green$'$s Function (NEGF) method in the multi-scale \cite{chen2015nemo5} and multi-physics \cite{miao2016buttiker, chen2015surface} Nano-Electronic MOdeling (NEMO5) tool  \cite{fonseca2013efficient}. 

The default parameters of the TE-TFET device ($L_{ch}$, $L_{ext}$ and $T_{ext}$ shown in Fig.~\ref{1}a) are set to $12nm$, $4nm$ and 3L respectively. $V_{DS}$ is 0.4V for $L_{ch}$ 12nm. Source and drain regions are doped with the doping level of $10^{20} cm^{-3}$. Equivalent oxide thickness (EOT) is set to 0.5nm. Constant field scaling E $ = 30V/nm$ is chosen for the device scaling, where $E = V_{DD}/L_{ch}$. 


\section{Results and Discussion}

\begin{figure}[!t]
\centering
{\includegraphics[width=0.36\textwidth]{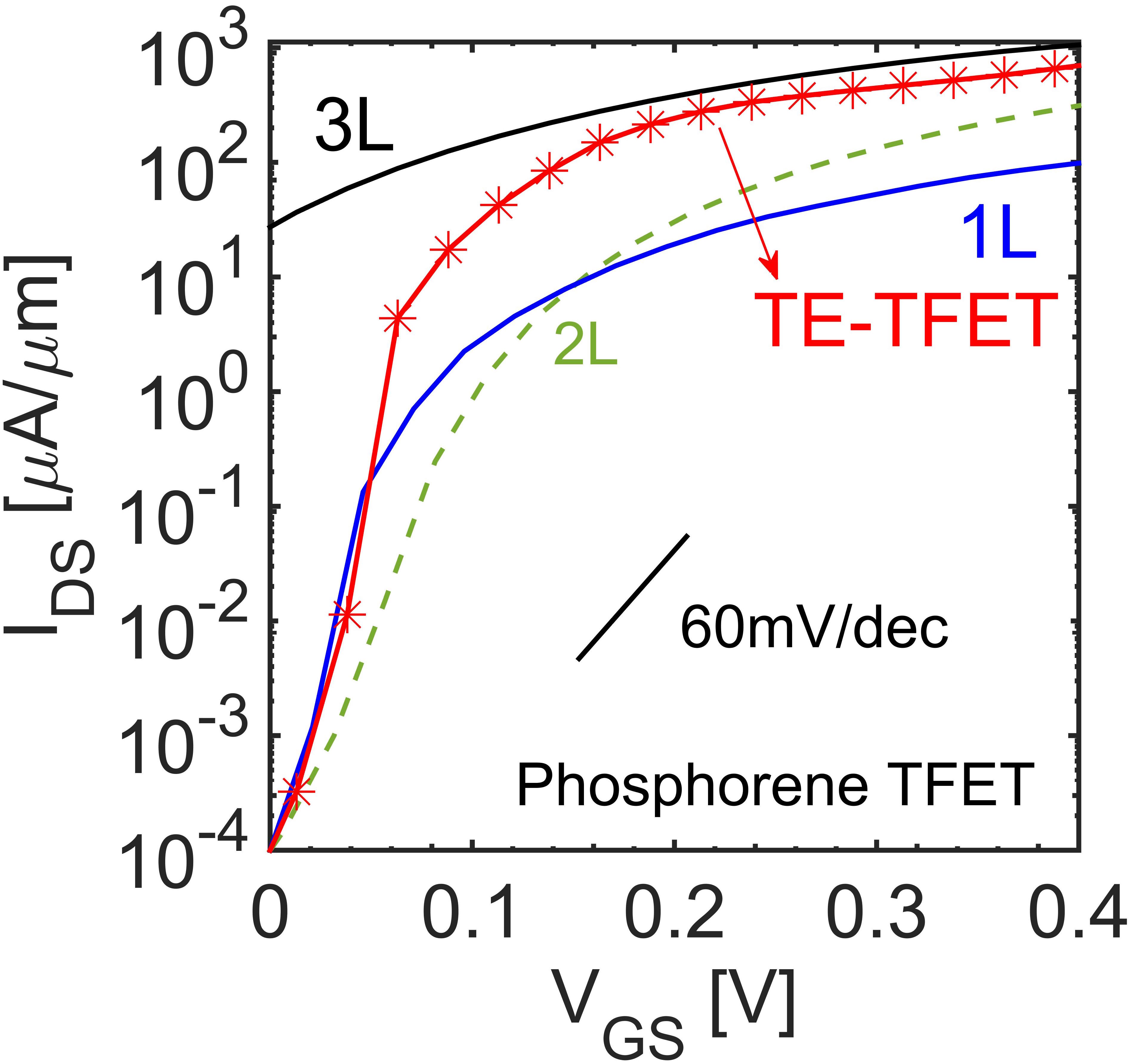}%
\label{20}}
\caption{The transfer characteristics of TE-TFET compared with different layer thickness TFET based on phosphorene. }
\label{2}
\end{figure}

The transfer characteristics ($Id-Vg$) of TE-TFET compared against 1L, 2L, and 3L phosphorene TFETs is shown in Fig.~\ref{2}. All the $Id-Vg$ curves, except the 3L-TFET, are shifted to have the same $I_{OFF}$ of $ 10^{-4} \mu A/\mu m$. The minimum current of 3L-TFET is $4.131\mu A/\mu m$, which is larger than the required $I_{OFF}$ level. Hence, 3L phosphorene TFET is shifted by the same voltage shift as that of the TE-TFET. 

Fig.~\ref{2} shows that TE-TFET has the advantages of both 3L and 1L homojunction TFETs: small $I_{OFF}$ of 1L-TFET and high ON-current of 3L-TFET. The $I_{ON}$ of about $700uA/um$ is achieved in TE-TFET with $V_{DD} = 0.4V$, which is 2 times larger than 2L-TFET.  $I_{60}$, the current when SS becomes $60mV/dec$ \cite{7021205}, in TE-TFET is about $10\mu A/\mu m$ which is larger by two orders compared to that of the best phosphorene TFET. 

\begin{figure}[!t]
\centering
\subfloat[]{\includegraphics[width=0.217\textwidth]{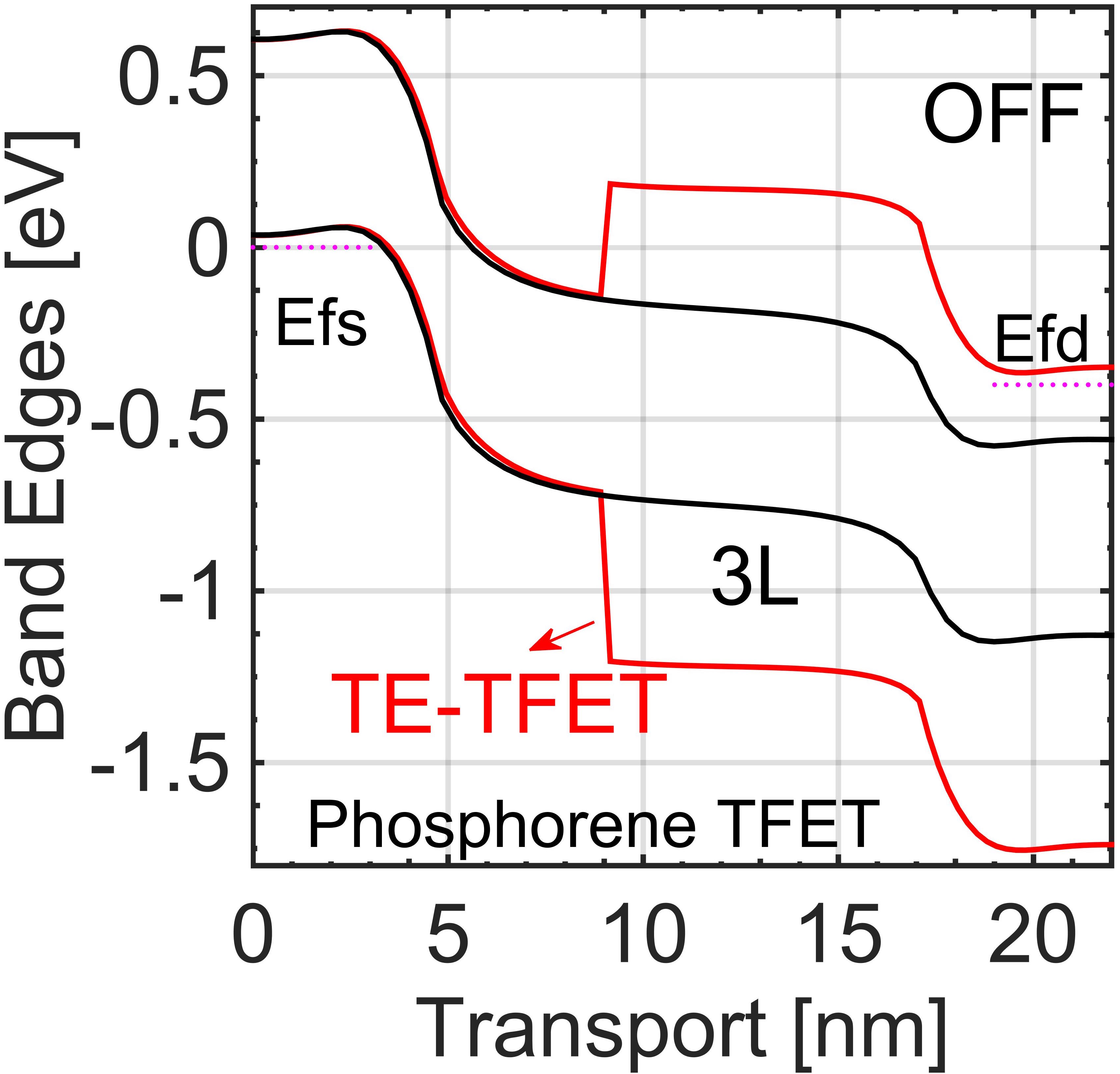}%
\label{2a}}
\hfil
\subfloat[]{\includegraphics[width=0.225\textwidth]{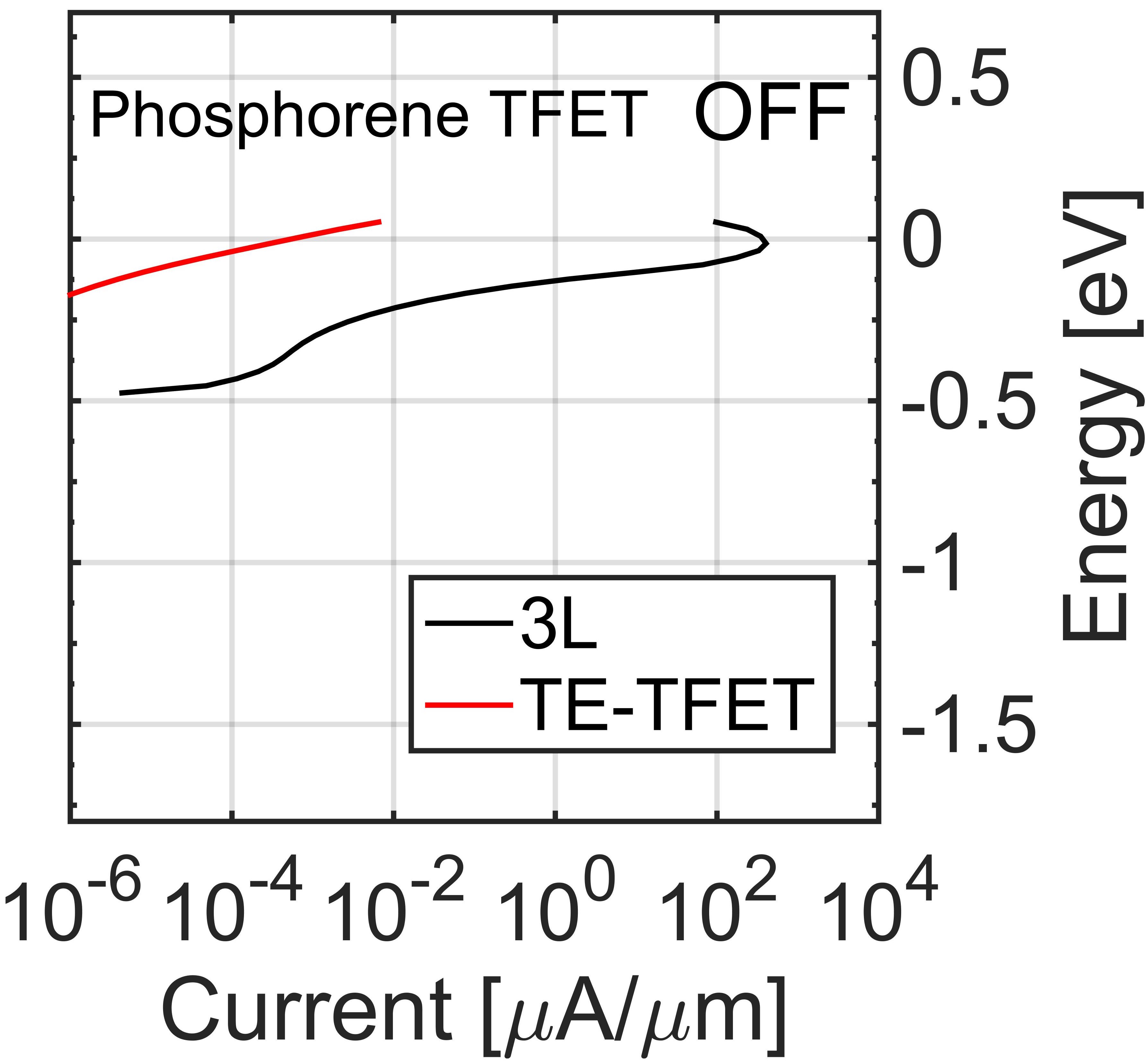}%
\label{2b}}
\vfil
\subfloat[]{\includegraphics[width=0.217\textwidth]{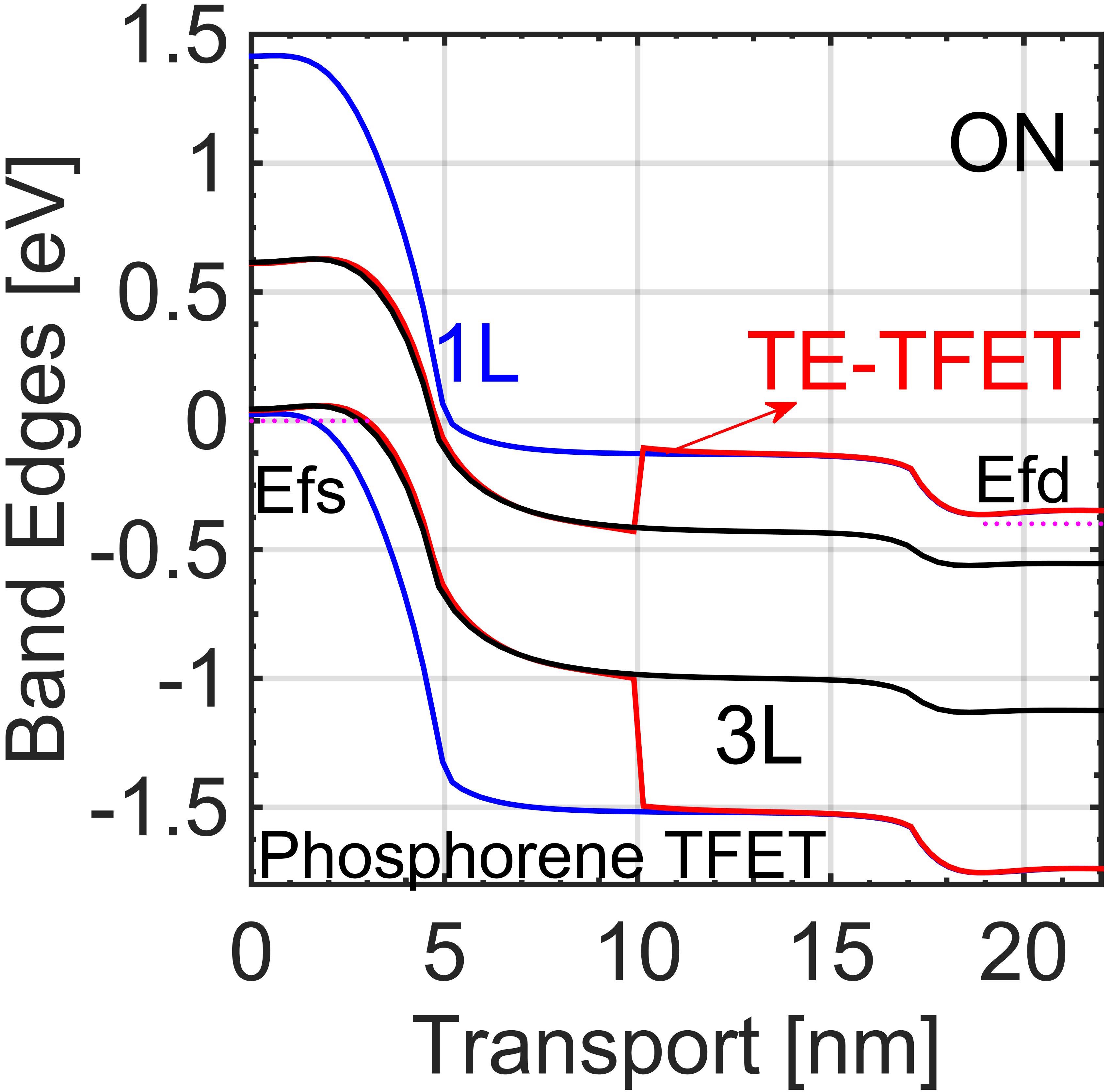}%
\label{2c}}
\hfil
\subfloat[]{\includegraphics[width=0.228\textwidth]{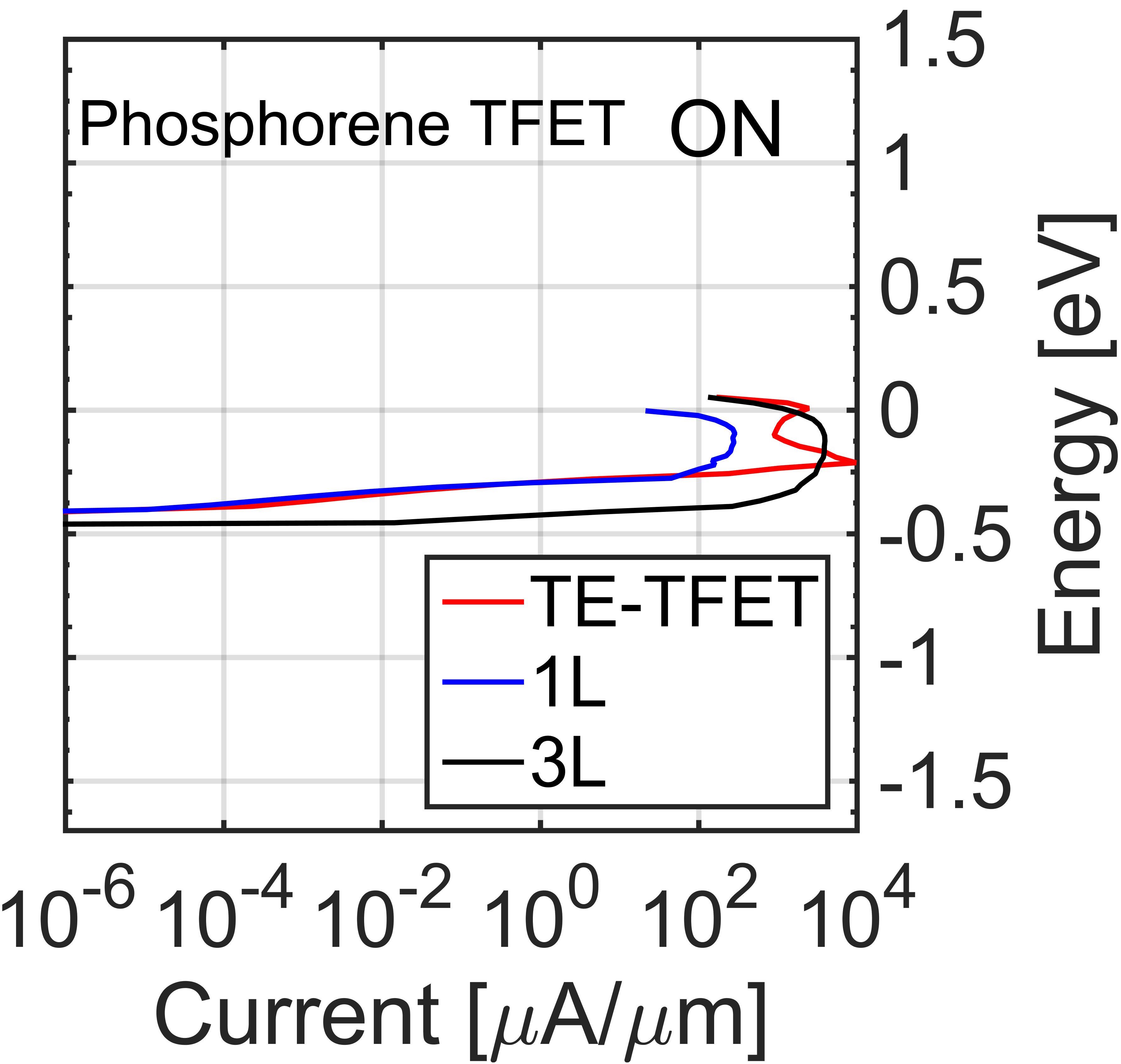}%
\label{2d}}
\caption{The band edges of (a) TE-TFET with 3L TFET at OFF state and (b) TE-TFET with 1L and 3L TFET at ON state aligned with the energy resolved current (c) and (d) respectively.}
\label{22}
\end{figure}

A comparison of TE-TFET against 3L-TFET in the ON-state and 1L-TFET in the OFF-state are shown in Fig.~\ref{22}; the band diagrams along with energy resolved current of these devices are also plotted. At OFF state, as shown in Fig.~\ref{22}a and b, TE-TFET has a larger barrier compared to the 3L-TFET which blocks the direct source-to-drain tunneling in the 1L section. TE-TFET thus achieves a small OFF-current. In the ON-state, shown in Fig.~\ref{22}c and d, TE-TFET has a smaller tunnel distance compared to the 1L-TFET. Thus, TE-TFET is able to achieve a higher ON-current compared to 1L-TFET due to the smaller tunnel distance at the source-channel interface. TE-TFET has $SS = 15mV/dec$ over four decades of drain current. In spite of high current levels in TE-TFET, its ON-current does not reach that of 3L-TFET due to the 1L barrier inside channel that blocks the current as shown in Fig.~\ref{22}c and d.\\

\begin{figure}[!t]
\centering
\subfloat[]{\includegraphics[width=0.23\textwidth]{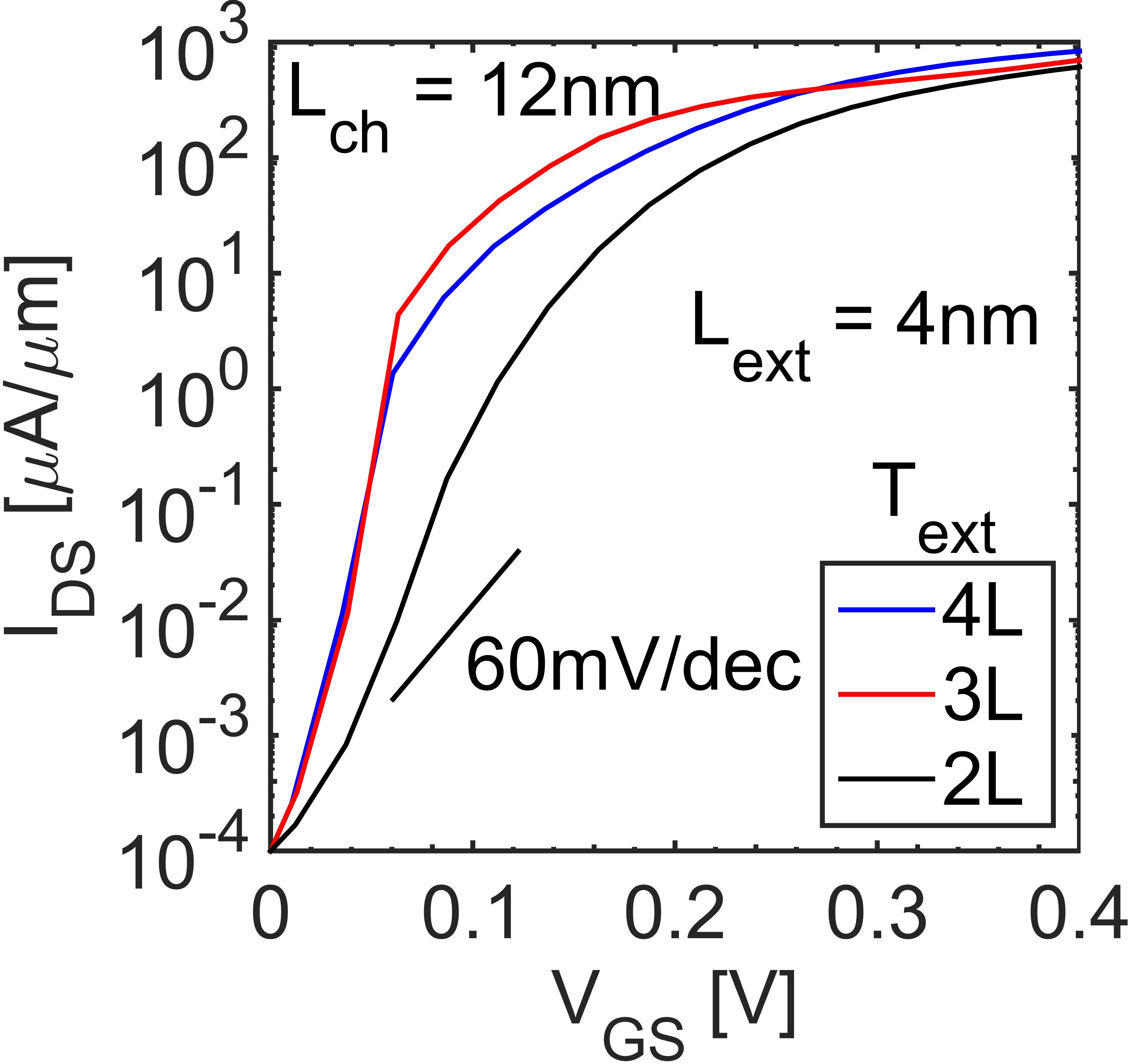}%
\label{4a}}
\hfil
\subfloat[]{\includegraphics[width=0.23\textwidth]{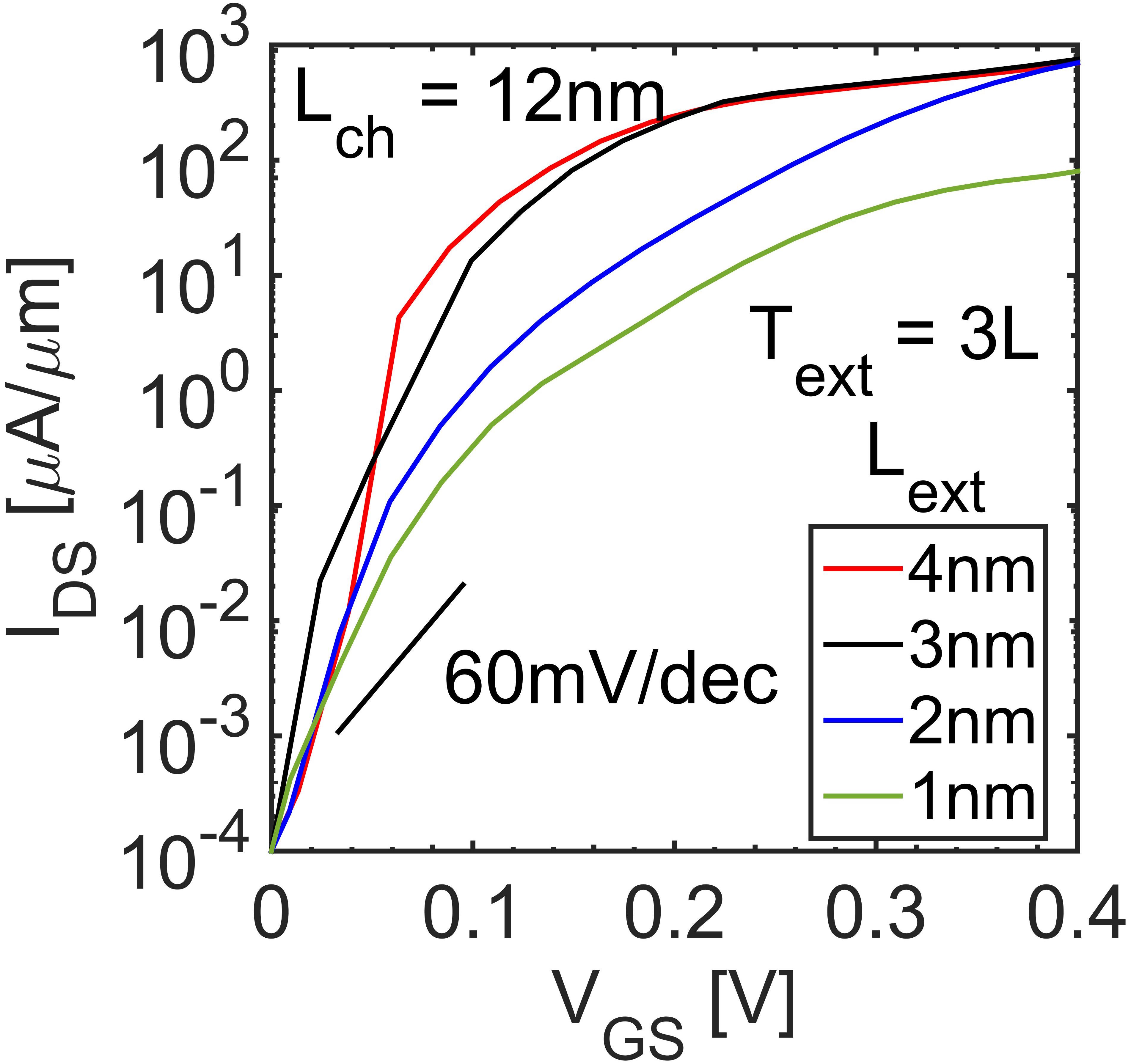}%
\label{4b}}
\vfil
\subfloat[]{\includegraphics[width=0.23\textwidth]{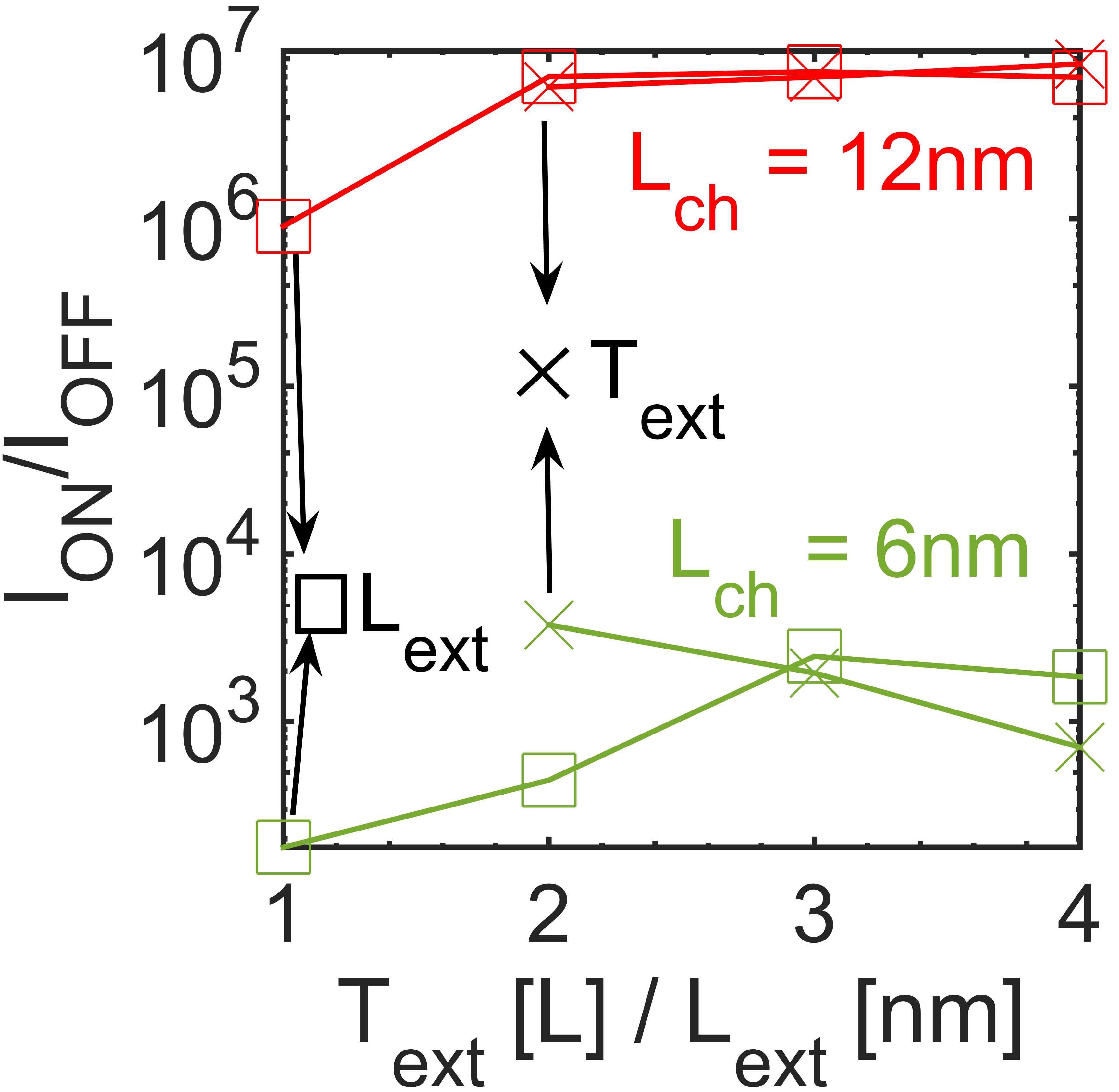}%
\label{4c}}
\hfil
\subfloat[]{\includegraphics[width=0.23\textwidth]{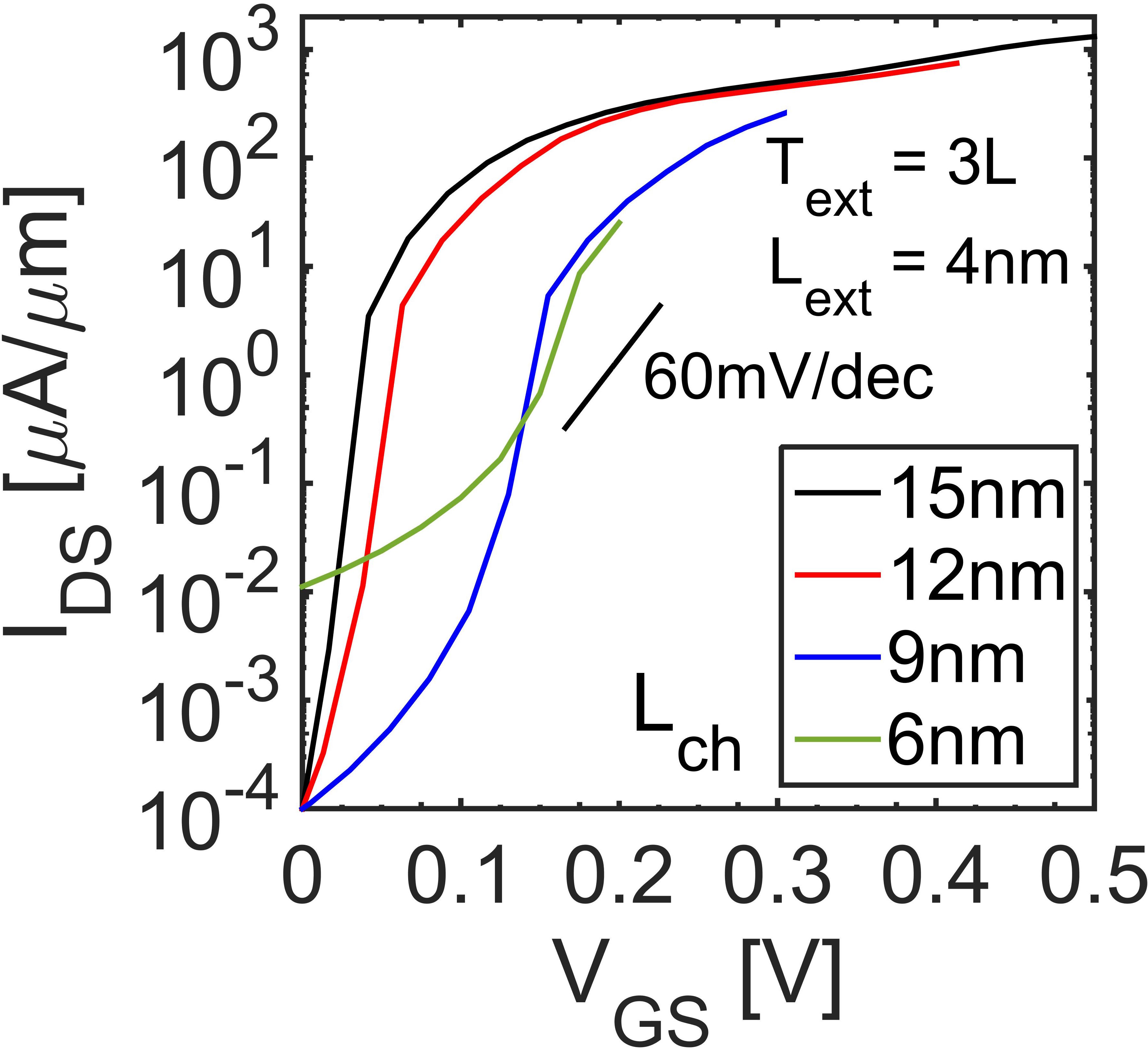}%
\label{4d}}
\caption{Id-Vg curves for TE-TFETs of $L_{ch} = 12nm$ with different (a) $T_{ext}$ and (b) $L_{ext}$. (c) The $I_{ON}/I_{OFF}$ change with respect to $T_{ext}$ and $L_{ext}$ for $L_{ch} = 12nm$ and $6nm$. (d)The transfer characteristics with $L_{ch}$ of TE-TFETs with constant field scaling ($E = V_{DD}/L_{ch}$ = 30V/nm) from 15nm to 6nm.}
\label{4}
\end{figure}

The impact of device design parameters $T_{ext}$ and $L_{ext}$ on its performance is discussed for a TE-TFET with $L_{ch} = 12nm$. As shown in Fig.~\ref{4}a, 4L, 3L and 2L flakes are used in the extension region of TE-TFET which translate into a bandgap of 0.481eV, 0.570eV and 0.803eV respectively.  The ON-current can be improved from $700 \mu A/\mu m$ to $800 \mu A/\mu m$ by replacing 3L with 4L in the extension region. TE-TFET with 2L extension still achieves an ON-current similar to 3L, but its $I_{60}$ degrades by two orders of magnitude. Fig.~\ref{4}b shows the impact of $L_{ext}$ on the performance of TE-TFET; by increasing $L_{ext}$ from 1nm to 2nm, the ON current improves by an order of magnitude. However, the performance saturates for $L_{ext}$ beyond 2nm (up to 4nm). This minimum value of $L_{ext}$ is because for the cases where $L_{ext}$ is too short, the tunneling in the ON-state does not occur completely in the small Eg region. Hence, a lower ON-current is achieved with $L_{ext}$ below 2nm.

The $I_{ON}/I_{OFF}$ ratio as a function of $T_{ext}$ and $L_{ext}$ for $L_{ch} = 12nm$ and $6nm$ is plotted in Fig.~\ref{4}c. For the 6nm channel length with $V_{DS}$ of 0.2V (constant field scaling), the trend is similar to that of $L_{ch} = 12nm$. The OFF current of TE-TFET with $L_{ch} = 6nm$ increases beyond $10^{-4}\mu A/\mu m$ due to the fact that 1L barrier inside the channel is not long enough to block the leakage current. For a fair comparison, $I_{ON}$ is fixed to $10^{2} \mu A/\mu m$ for $T_{ext}$ analysis, whereas $I_{OFF}$ is set to $10^{-3} \mu A/\mu m$ for $L_{ext}$ study. It is worthwhile to mention that there is only a small range of $V_{GS}$ for $L_{ch} = 6nm$ in which the SS is smaller than $60mV/dec$. \\

\begin{figure}
\centering
\subfloat[]{\includegraphics[width=0.23\textwidth]{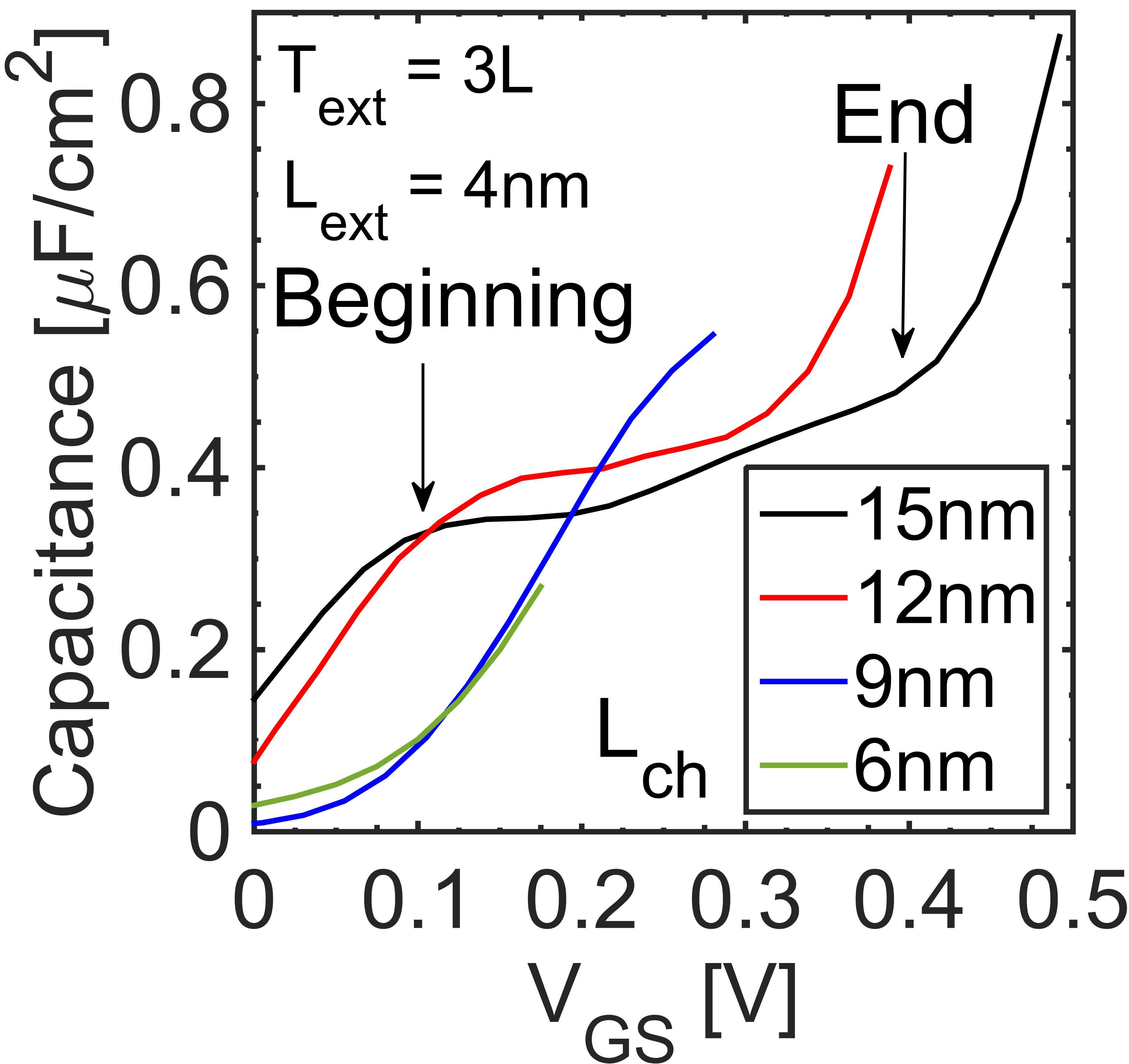}%
\label{5a}}
\hfil
\subfloat[]{\includegraphics[width=0.23\textwidth]{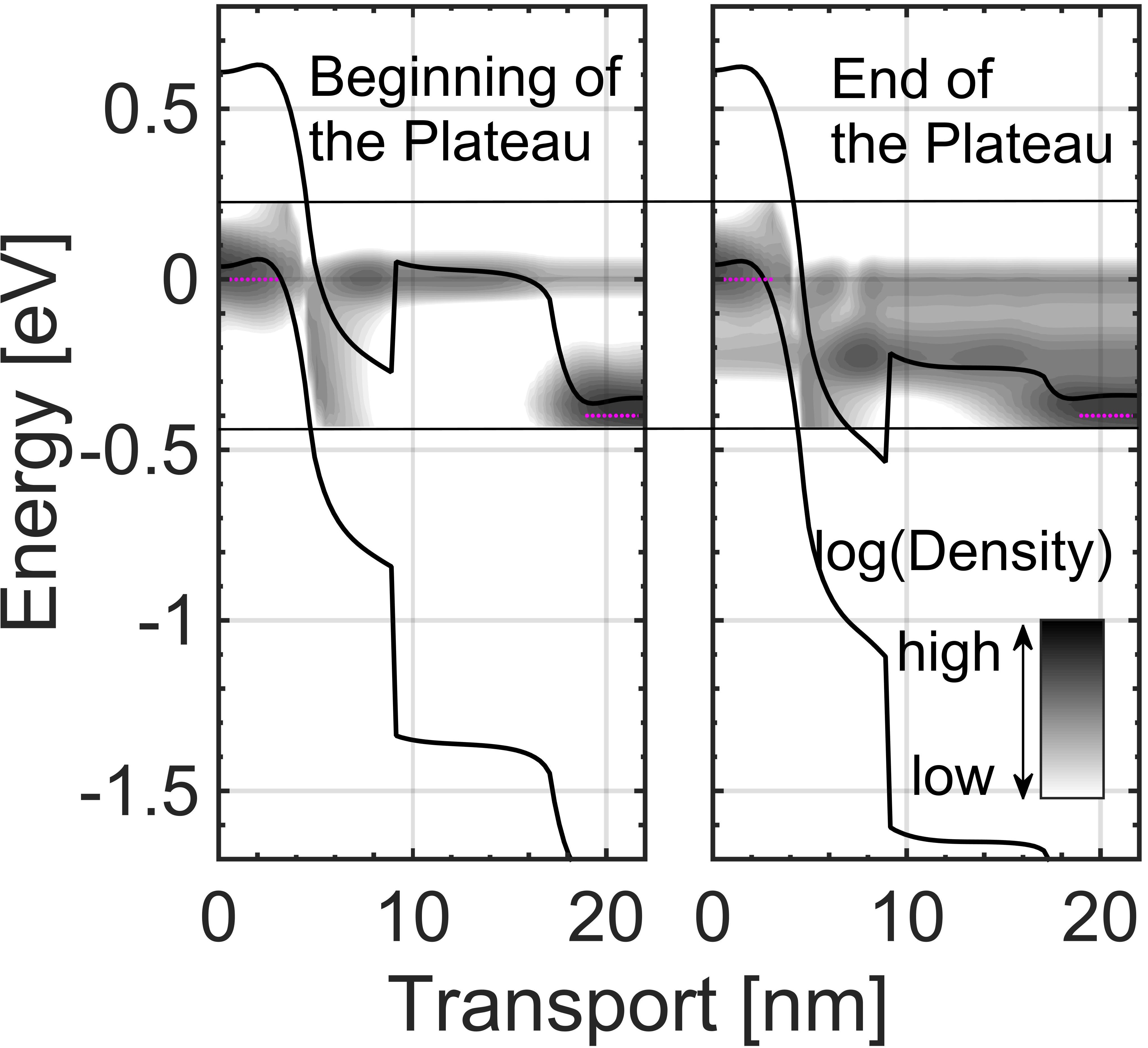}%
\label{5b}}
\caption{(a) The transfer characteristics and (b) C-V of with $L_{ch}$ of TE-TFETs with constant field scaling ($E = V_{DD}/L_{ch}$ = 30V/nm) from 15nm to 6nm.}
\label{5}
\end{figure}

Constant field scaling E  $ = 30V/nm$ of TE-TFETs is studied in this part ($E = V_{DD}/L_{ch}$). Fig.~\ref{4}d shows that the $I_{ON}/I_{OFF}$ of TE-TFET is over 6 orders of magnitude for $L_{ch}$ above 9nm, however there is a noticeable increase in $I_{OFF}$ for the channel length of 6nm.  Fig.~\ref{5}a shows the impact of scaling on the total gate capacitance characteristics ($C-V_{GS}$). The gate capacitance is noticeably smaller than most TMD materials. This smaller capacitance originates from the smaller effective mass of phosphorene \cite{ameen2015few}. Unlike homojunctions, the CV curve of TE-TFET has a plateau region. This plateau in CV appears due to the strong density of states (DOS) modulation within the quantum well region. Fig.~\ref{5}b illustrates the carrier density along the transport direction at the beginning and the end of the plateau region.  In both cases, source Fermi level is aligned with the maximum DOS. From the beginning to the end, $Ef_s$ is aligned with a lower DOS due to the confinement. This decrease in DOS in the quantum well region compensates for the increase in DOS inside the 1L region and forms the plateau in the C-V curves. The length of the plateau region is different for different $L_{ch}$ since the carrier density is also influenced by the carrier injection from drain \cite{7336792}.

\begin{figure}[!t]
\centering
{\includegraphics[width=0.36\textwidth]{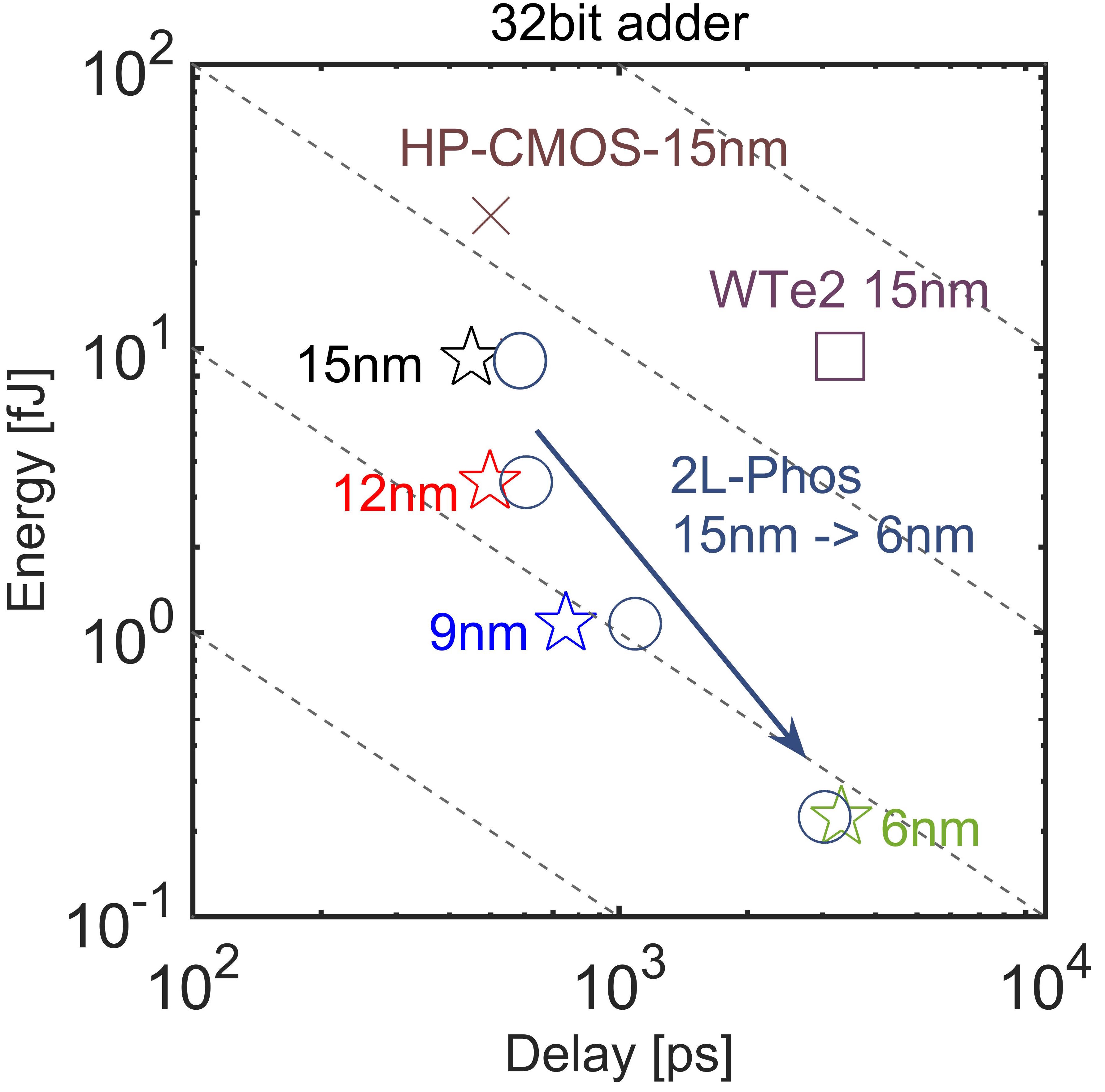}%
\label{60}}
\caption{Energy-Delay product of TE-TFETs in comparison with 2L-TFETs \cite{ameen2015few} for $L_{ch}$ from 15nm to 6nm and a 15nm WTe$_2$ TFET. }
\label{6}
\end{figure}

Compared to homojunction phosphorenene TFETs \cite{ameen2015few}, TE-TFETs exhibit higher ON-currents and slightly higher capacitances. These higher ON-currents translate to an improvement in 32 bit adder energy-delay product as shown in Fig.~\ref{6}. The 32-bit adder energy-delay product is calculated using BCB 3.0 model \cite{18} in which the parasitic capacitances are taken into account. The circuit parameters required in BCB model are taken from ITRS roadmap. 

\section{Conclusion}
In conclusion, thickness engineered tunneling field-effect transistor (TE-TFET)is proposed and evaluated in this work. By taking advantage of flake-thickness-dependent direct bandgap in phosphorene, an artificial heterostructure TFET can be achieved. The absence of interface, between different materials in artificial heterojunctions, allows TE-TFET to avoid the interface states and lattice mismatch problems observed in conventional heterojunction TFETs while providing similar boost in the ON-current of 1280$\mu A/\mu m$ with a 15nm channel length. TE-TFETs are scalable down to 9nm with constant field scaling  E $ = V_{DD}/L_{ch}= 30V/nm$. Offering higher ON-current, TE-TFETs outperform the best homojunction phosphorene TFETs and TMD TFETs in terms of circuit energy-delay product. 

\begin{acknowledgments}
This work was supported in part by the Center for Low Energy Systems Technology, one of six centers of STARnet, and in part by the Semiconductor Research Corporation Program through Microelectronics Advanced Research Corporation and Defense Advanced Research Projects Agency.\\
\end{acknowledgments}

\bibliography{all.bib}

\end{document}